# Evidence of nematic order and nodal superconducting gap along [110] direction in RbFe$_2$As$_2$

Xi Liu[1†], Ran Tao[1†], Mingqiang Ren[1†], Wei Chen[1], Qi Yao[1], Thomas Wolf[3], Yajun Yan[1], Tong Zhang[1,2*], Donglai Feng[1,2*]

[1]State Key Laboratory of Surface Physics, Department of Physics, and Advanced Materials Laboratory, Fudan University, Shanghai, 200433, China

[2]Collaborative Innovation Center of Advanced Microstructures, Nanjing, 210093, China

[3]Institute for Solid State Physics, Karlsruhe Institute of Technology, D-76021 Karlsruhe, Germany

† These authors contributed equally to this work.
*Email: tzhang18@fudan.edu.cn, dlfeng@fudan.edu.cn

**Unconventional superconductivity often intertwines with various forms of order, such as the "nematic" order which breaks the rotational symmetry of the lattice. Investigation of these ordered phases sheds crucial light on the superconductivity itself. Here we report a low-temperature scanning tunneling microscopy (STM) study on RbFe$_2$As$_2$, a heavily hole-doped Fe-based superconductor (FeSC). We observe significant symmetry breaking in its electronic structure and magnetic vortex which differentiates the (π, π) and (π, -π) directions of the unfolded Brillouin zone (BZ). It is thus a novel nematic state, distinct from the nematicity of undoped/lightly-doped FeSCs which breaks the (π, 0) / (0, π) equivalence. Moreover, we observe a clear "V"-shaped superconducting gap which can be well fitted with a nodal gap function. The gap is found to be suppressed on surface Rb vacancies and at step edges, and the suppression is particularly strong at the [110]-oriented edges. This is possibly due to a d$_{x^2-y^2}$ like pairing component with nodes along the [110] directions. We further demonstrated that such (π, π) nematic state can be suppressed via surface electron doping to RbFe$_2$As$_2$, and the superconductivity is subsequently enhanced. Our results thus highlight the intimate connection between nematicity and superconducting pairing in iron-based superconductors.**

The discovery of FeSCs has opened a new era in the study of unconventional superconductivity[1-3]. Most FeSCs are found to be proximate to a magnetically ordered state and a nematic electronic state that shares similarities with the cuprates[4-12]. In most undoped and lightly-doped FeSCs, the Fe ions are close to 3d$^6$ configuration which favor a stripe-like collinear antiferromagnetic (AFM) order or spin density wave (SDW), with a wave vector Q = (π, 0) or (0, π) (an exception is FeTe, which has an bicollinear AFM state with Q = (π/2, π/2)). A nematic phase which breaks the equivalence between *a* and *b* directions in the Fe-plane develops at the Neel temperature ($T_N$) or slightly above, and approaches the superconducting dome upon doping[4-12]. There has been increasing evidence showing the nematicity is driven by magnetic fluctuations[13-14], nonetheless the orbital-driven scenario is also proposed, especially for FeSe (ref. 15). The magnetic and/or orbital fluctuations between the nested Fermi surfaces with a vector around Q may play an essential role in superconductivity and determine the pairing symmetry[1-4]. Therefore, the relation between nematicity, magnetic order and superconductivity has become one of the central themes in FeSCs.

To provide a unified understanding of FeSCs and even cuprates, it is important to examine such a theme in regimes where the configuration of Fe ions deviates significantly from $3d^6$, as the magnetic interactions, electron correlations and Fermi surface topology will alter drastically[2,4,16]. In theory, strong correlations are expected for the $3d^5$ case to drive the system into a Mott insulating phase[17,18]. It has been suggested that electronic nematicity may occur via doping a Mott insulator[19], as evidenced in underdoped cuprates[20,21]. Meanwhile, the pairing symmetry of FeSC is also predicted to vary with doping[2,3,16]. For the $3d^{5.5}$ configuration, the dominant spin fluctuations are predicted to relocate to $(\pi, \pi)/(\pi, -\pi)$ due to a change in Fermi surface topology, and consequently, $d$-wave pairing is favored[16,22,23]. This configuration has been realized in $AFe_2As_2$ (A=K, Rb, Cs), the most heavily hole-doped FeSCs[24-41]. They have large Sommerfeld coefficients ($\gamma$)[24,25] and mass enhancement[27,33,35], indicative of strong correlations. Recently, a coherence-incoherence crossover[26] and heavy-fermion like behavior[28] were observed in $AFe_2As_2$, suggesting an orbital-selective Mott transition. Heat transport[29,31], magnetic penetration depth[30], NMR[41] and ARPES[34,36] measurements have suggested gap nodes in $AFe_2As_2$ (A=K, Rb); however, whether the nodes are symmetry protected (d-wave pairing)[16,22] or accidental (from anisotropic s-wave pairing)[42,43] remains hotly contested. In parallel, neutron scattering[39,40] and NMR[41] studies on $KFe_2As_2$ did reveal spin fluctuations that deviated from $(\pi, 0)$. It is thus critical to look for possible nematicity with distinct behaviors and its relation with superconductivity.

In this article, we present a milliKelvin STM study on $RbFe_2As_2$ single crystals. Compared to its sister compound $KFe_2As_2$, $RbFe_2As_2$ has an even larger $\gamma$ value (~127 mJ/mol·K$^2$ (ref. 25)) but a lower $T_c$ (~2.5K). Remarkably, we observed significant twofold symmetry in the quasi-particle interference (QPI) and magnetic vortex cores, while the surface atomic lattice remains four-fold symmetric within the experimental resolution. Particularly, this $C_4$ - $C_2$ symmetry breaking is along the diagonal direction, 45° off from the Fe-Fe bond or the $(\pi, \pi)$ direction in the unfolded BZ. This suggests that a new type of electronic nematicity and associated fluctuations developed in $RbFe_2As_2$, and such diagonal nematicity was found to persist above Tc. Moreover, high energy-resolved tunneling spectra revealed a clear "V"-shaped superconducting gap which can be well fitted by a nodal gap function. The gap is suppressed by both surface Rb vacancies (non-magnetic impurities) and near atomic step edges, suggestive of sign change pairing. Moreover, the spatial extension of the suppressed-gap region on [110] oriented edges is found to be much wider than the [100] oriented edges, which is likely due to gap nodes in the [110] directions. Finally, we performed surface K dosing on $RbFe_2As_2$ and demonstrate that the $(\pi, \pi)$ nematic state can be suppressed by electron doping, while the superconductivity is subsequently enhanced. The possible origin of the diagonal nematicity and its relation to superconducting pairing is discussed.

**Results:**

**Surface atomic structure and superconducting gap of $RbFe_2As_2$:**

The experiment was mostly conducted in a millikelvin STM working at $T$ = 20mK (the surface K-dosing was conducted in a 4.5K STM system). The effective electron temperature ($T_{eff}$) of the former system is calibrated to be 310 mK (see section II of the Supplementary Material). Sample preparation and more experimental details are described in **Methods**. $RbFe_2As_2$ is stoichiometric with the $ThCr_2Si_2$-type structure (Fig. S1**a**). It is expected to cleave between FeAs layers and results Rb covered surfaces. Fig. 1a shows the typical topography of a commonly observed surface (referred as type A surface). It is atomically flat with some basin-like defects. In the defect-free areas, a square lattice with an inter-atomic spacing of 5.4 Å is observed (Fig. 1a inset). This spacing is $\sqrt{2}$ times the in-plane lattice constant of RbFe$_2$As$_2$ ($a_0$ =3.86Å). Besides type A surface, we occasionally observed another type of surface region (type B surface), as shown in Fig. 1b. The topography of type B surface is actually similar to type A, and a 5.4 Å square lattice is also observed in its defect-free areas (Fig. 1b inset). However it displays much stronger anisotropy in the electronic states, as we will show below.

To determine the surface atomic structure, higher-resolution STM imaging near the basin defect is shown in Fig. 1c. We found that there is another square lattice *inside* the basin defect, with a lattice constant equal to $a_0$ (as marked by black dots). The outside lattice (marked by circles) is rotated 45° and forms a $\sqrt{2} \times \sqrt{2}$ reconstruction with respect to the inside lattice (see section III of the Supplementary Materials for more details). The surface structure that best explains these observations is sketched in Fig.1d: the surface is Rb terminated with 50% coverage, and the basin areas are Rb vacancies with an exposed FeAs layer. Surface Rb atoms are surrounded by four As atoms underneath, and formed a $\sqrt{2} \times \sqrt{2}$ lattice with the same orientation as the Fe lattice (as denoted by *a, b* hereafter). We note in this model the Rb lattice will have two equivalent occupation site that shifted by 1/2 lattice spacing (see Fig. S3e). We indeed observed domain structures formed by these two occupations on low-temperature cleaved sample (see Fig. S4). And the orientation of the surface Rb lattice is further confirmed by Laue diffraction combined with STM imaging (Fig. S5). Moreover, such a surface structure retains fourfold rotational symmetry, at least within the spatial resolution of STM, and is *non-polar* due to the 50% Rb coverage, which allows STM to access the intrinsic electronic states of RbFe$_2$As$_2$. A similar surface structure was also observed on cleaved KFe$_2$As$_2$ (ref. 38).

Fig.1e shows typical tunneling spectra on defect-free areas of type A and B surfaces, within a relatively large energy scale (±200 mV). On both regions a pronounced conductance peak is observed slightly below $E_F$ (at about -2 ~ -3 meV). The peak is asymmetric with a higher-intensity shoulder at negative energy. This is likely due to a hole-like band with a top just below $E_F$, as evidenced in the QPI measurements below. The enhanced density of states (DOS) near $E_F$ may underlie the large $\gamma$ value of RbFe$_2$As$_2$. A similar conductance peak near $E_F$ was also observed in KFe$_2$As$_2$ (ref. 38).

To explore the superconducting state, low-energy tunneling spectra (±1.6mV) were measured at zero magnetic field. As shown in Fig. 1f, a well-defined "V"-shaped superconducting gap was observed on type A surface (blue dots), which is the commonly observed case; for type B surface the gap is noticeably broader (green dots). The hump-like structure at negative bias is due to the aforementioned strong DOS peak. The gap on both surfaces are spatially uniform (see Fig. S6). We found that the superconducting gap of type A surface can be well fitted with a nodal gap function. The red curve in Fig. 1f is a d-wave fit by using the Dynes formula[44] for the superconducting DOS:

$$N(E)_k = |Re[(E - i\Gamma)/\sqrt{(E - i\Gamma)^2 - \Delta_k^2}]|, \quad \text{and} \quad \Delta_k = \Delta_0 \cos(2\theta_k)$$

(Note that the gap function $\Delta_k = \Delta_0 \cos(4\theta_k)$ which corresponds to nodal s-wave pairing[42] will result in exactly the same DOS). The tunneling conductance is given by $dI/dV \propto \int N(E)_k f'(E + eV) dk dE$, where $f(E)$ is the Fermi-Dirac function at $T_{eff}$ = 310 mK. The fitting yields $\Delta_0$ = 0.47 meV and a small $\Gamma$ of 0.03 meV that accounts for additional non-thermal broadening (*e.g.* impurity scattering). The ratio $2\Delta_0/k_B T_c$ is 4.36. For comparison, an isotropic

*s*-wave gap fit is also plotted in Fig. 1**f** (dashed curve). It does not match the tunneling spectrum, especially around the gap bottom. On type B surface, the nodal fit with a similar gap size of $\Delta_0$ = 0.46 meV can match the gap bottom but deviates near the coherence peaks (black curve). The fitting yields a larger $\Gamma$ of 0.09 meV which may indicate a detrimental effect to superconductivity on type B surface.

**$C_4$ symmetry breaking in QPI and magnet vortex mapping:**

Next we turn to examine the electronic structure of RbFe$_2$As$_2$ by performing dI/dV mapping. Fig. 2**a** shows two representative dI/dV maps taken on the type A surface in Fig. 1**a** (at T=20mK, $T_{eff}$ =310 mK), where clear interference patterns can be observed. Fig. 2**c** displays the raw and symmetrized fast-Fourier transform (FFT) maps at different energies, which give the *q*-space scattering patterns (see Fig. S7 for a complete set of QPI data and raw FFTs). The FFTs display a ring-like structure at relatively high energies ($E > 7$ meV). However, it becomes diamond-shaped and obviously twofold symmetric as approaching $E_F$. To identify the orientation of the scattering pattern, in Fig. 2**e** we plot the FFT map at $E$ = 2.2 meV together with the unfolded BZ (derived from atomic resolved topography and the lattice structure shown in Fig. 1**d**). It is seen that the twofold-symmetric axes are along the (π, π) or (π, -π) directions (the *diagonal* of the Fe plaquette). Such $C_4$ symmetry breaking between (π, π) and (π, -π) in QPI has never been reported for other iron pnictides. Another notable feature is the elongated direction of the scattering pattern rotates 90° as the energy is lowered below $E_F$ (e.g. compare the $E$ = 2.2 meV and -2.7meV panel in Fig. 2**c**, and see also Fig. S7). Despite such a complicated evolution, the overall anisotropic scattering patterns changes with energy (see also Fig. S7), indicating that they originate from QPI of anisotropic band(s). To demonstrate the evolvement of such band(s), in Fig. 2**f** we summarized the FFT profile near (π, 0) direction (where strong scattering weight distributed around) at various energy. An overall hole-like dispersion can be seen, and a parabolic fit yields a band top ($E_b$) = 27 meV and Fermi crossing at $q_F$ = 0.21 Å$^{-1}$. More specific interpretation to such QPI will require detailed knowledge on the origin of anisotropic band structure.

Besides commonly observed type A surfaces, on the occasionally observed type B surfaces we found even greater anisotropy. Figs. 2**b,** 2**d** show representative dI/dV maps and FFTs taken on a 50 × 50 nm$^2$ area marked in Fig. 1**b** (a complete set of QPI data is shown in Fig. S8). Highly anisotropic interference patterns can be seen in the vicinity of surface defects, which are more pronounced along one of diagonal direction of the Fe lattice. The corresponding FFTs now display two arc-like features which are also along the (π, π) in the unfolded BZ, manifesting a strong $C_4$ symmetry breaking. Such arc-like features exist in a relatively wide energy range of -10meV ~ 20meV (see Fig. S8) and disperse with energy as well. In Fig. 2**g** we show the FFT profile of type B surfaces around the (π, π) direction. A hole-like dispersion is observed with $E_b$ = 30 meV and $q_F$ = 0.22 Å$^{-1}$, which are close to the values for type A surface. Thus the basic band structure of type B surface is likely similar to that of type A surface; however, it was driven to be more anisotropic for some reason as discussed later. Despite the anisotropy, the hole-like dispersion in QPI appears to be consistent with a recent ARPES study on RbFe$_2$As$_2$ (ref. 37), in which a single hole pocket is observed at $\Gamma$. We note that ARPES studies on the sister compound KFe$_2$As$_2$ found three hole pockets at $\Gamma$, which is reproduced in DFT calculations on its paramagnetic state[33-35]. As discussed in Ref. 37, this difference could be due to the larger spacing between FeAs layers in RbFe$_2$As$_2$, which enhances the two-dimensionality of the system.

Another notable feature in Fig. 2**f** is that besides the main hole-like dispersion, there is likely a second band with the top very close to $E_F$ (tracked by black dashed line). Such feature is more clearly seen in the FFT profiles of type B surface along the (π, -π) direction, as shown in Fig. 2**h**. However it is totally absent in the (π, π) direction (Fig. 2**g**), which also reflects the $C_4$ - $C_2$ symmetry breaking. This band may closely relate to the DOS peak observed in dI/dV (Fig. 1**e**), as there could be van-Hove singularity near the band top[38]. Its exact origin is unclear at this stage.

To further investigate this novel $C_4$ symmetry breaking and its effect on the superconductivity, we studied magnetic vortices induced by an external field. Fig. 3**a** shows a zero-bias conductance (ZBC) mapping taken on a 150 × 150 nm$^2$ area of type A surface, under a perpendicular field of $B$ = 0.5T. A vortex lattice is reflected by the high conductance regions. One sees that the vortex cores display anisotropic shape. To show this more clearly, a spatially averaged core (of 6 vortices) is shown in the Fig. 3**c** inset. It is slightly elongated along the diagonal of the Fe lattice, consistent with the twofold symmetry of QPI. At the core center, a zero-bias peak is observed in dI/dV (Fig. 3**b**), and the peak "splits" on moving away from the center. This is typical behavior of vortex core states for a clean superconductor[45]. The core states decay spatially on approximately the scale of the superconducting coherence length ($\xi$). Exponential fits to the profile along the long and short axes of the vortex core yield $\xi_A^L$ =15.0(± 0.35) nm and $\xi_A^S$ =12.5(± 0.2) nm (Fig. 3**c**), respectively. For type B surface, ZBC map of a 225 × 225 nm$^2$ area under B = 0.5T is shown in Fig.3**d**. The vortex cores are clearly more anisotropic (a zoom-in of single core is shown in Fig. 3**f**); while a zero-bias peak is also observed in the core center (Fig. 3**e**). The coherence lengths for the long and short axes are found to be $\xi_B^L$ =26.7±0.7 nm and $\xi_B^S$ =16.3±0.5 nm, respectively. As expected, the ratio $\xi^L/\xi^S$ of type B surface (1.63) is larger than that of type A surface (1.2), reflecting stronger anisotropy in the former. Since the spatial shape of the vortex core is intimately related to the underlying band structure[46], the elongated vortex cores provide further evidence for $C_4$ symmetry breaking in the electron states. To see the orientation of the anisotropic core with respect to the **k**-space band structure, we superposed FFT maps (taken near $E_F$) onto the vortex maps as insets in Figs. 3**a** and 3**d**. On type B surface, the vortex is elongated along the direction where the FFT displays a (dispersive) arc-like feature. This is apparently consistent with the BCS expectation that $\xi$ is longer in the direction with larger Fermi velocity ($\xi \sim hv_F/\pi\Delta$). On type A surface, a similar tendency is observed, despite weaker anisotropy in the vortex core and QPI. We note the **k**-space structure of $\Delta$ and the possible "nematic order" as discussed below should also related to the anisotropy of vortex core[47,48].

So far, the $C_4$ symmetry breaking between (π, π) / (π, -π) has been clearly evidenced in QPI and vortex measurements. The emergence of significant symmetry breaking in such heavily hole-doped region is surprising. It cannot be a surface effect since the surface atomic structure remains four-fold symmetric (Fig. 1**d**), and there is no bulk structural transition reported for $AFe_2As_2$ (A= Rb, K, Cs). Furthermore, the surface Rb vacancies are point-like without noticeable uniaxial anisotropy that may introduce anisotropic QPI, and the shape of the vortex core is also unrelated to surface defects. Thus, the observed symmetry breaking is reminiscent of a nematic-like electron state. Previously, anisotropic QPI which breaks the symmetry between (π, 0) and (0, π) was observed in undoped and lightly-doped iron pnictides, such as $Ca(Fe_{1-x}Co_x)_2As_2$ (ref. 6), NaFeAs (ref. 11) and LaOFeAs (ref. 8). It was commonly considered to be a signature of electronic nematicity with an origin closely related to spin and/or orbital degrees of freedom. Theoretical works have shown that the (π, 0) stripe AFM (SDW) order tends to open a partial gap along the antiferromagnetic direction, which distorts the Fermi surface and thus the QPI to be twofold symmetric[49,50]. Such symmetry breaking may even persist above $T_N$ due to short-range spin fluctuations[51]. Here, it is the first time to visualize a $C_4$ symmetry breaking in heavily hole-doped FeSC, and in a 45° rotated direction, this may suggest RbFe$_2$As$_2$ is likely proximate to a stripe-type AFM order or SDW with a Q along ($\pi$, $\pi$) direction, which breaks the ($\pi$, $\pi$)/($\pi$, -$\pi$) equivalence. We note that twofold anisotropic QPI along ($\pi$, $\pi$) has been reported in FeTe films which exhibit a bi-collinear AFM with Q = ($\pi/2$, $\pi/2$) (ref. 52). While direct measurements of spin fluctuations in RbFe$_2$As$_2$ are still lacking, spin fluctuations that deviated from ($\pi$, 0) has been observed in KFe$_2$As$_2$ (refs. 39~41). However, whether these fluctuations can drive the $C_4$ symmetry-breaking needs further study. On the other hand, orbital order can also drive nematicity, as proposed for FeSe (ref. 15), which will result in an anisotropic Fermi surface and QPI[7]. A similar orbital order was also observed in FeSe$_x$Te$_{1-x}$ (ref. 53). No matter which mechanism may apply, details of the anisotropic band structure will depend on material parameters such as Hund's coupling and on-site Coulomb interaction[49,50], which require further investigations.

Assuming the symmetry breaking is from electronic nematicity, there remains a question as to why type B surface shows stronger anisotropy than type A surface, despite their seemingly identical surface lattice structure. We note that for undoped iron pnictides, the ($\pi$, 0) / (0, $\pi$) nematicity (and the stripe AFM order) can be enhanced by applying uniaxial pressure[54]. Thus we speculate that type B surface may have local strain (e.g. due to different shrinkage of the sample and its glue upon cooling), which enhances the nematic state in these regions, while even type A surface may also have strain but is likely weaker than type B surface. Nevertheless, even if the strain plays a role here, our observations still imply that RbFe$_2$As$_2$ has a strong tendency or "susceptibility" to form nematic states along ($\pi$, $\pi$) / ($\pi$, -$\pi$) rather than ($\pi$, 0) / (0, $\pi$). Furthermore, it is clear that the superconducting gap features observed on type B surface are significantly broader than on type A surface (Fig. 1**f**). This reflects a competition between superconductivity and the degree of anisotropy, resembling the anti-correlation between superconductivity and the ($\pi$, 0) / (0, $\pi$) nematicity observed in NaFe$_{1-x}$Co$_x$As (ref. 12) and FeSe$_x$Te$_{1-x}$ (ref. 53). More evidence on such anti-correlation behavior is shown in the surface K dosing measurement below.

**Effect of Rb vacancy and atomic step edges on superconductivity:**

The ($\pi$, $\pi$) / ($\pi$, -$\pi$) symmetry breaking may have a more profound relation to the superconductivity in RbFe$_2$As$_2$ – the interactions and fluctuations of the nematic state could also underlie the electron pairing. Investigating the pairing symmetry of the system can provide further insight on this. In Fig. 1**f**, the well-defined 'V'-shaped superconducting gap has suggested a nodal pairing, to further investigate the pairing symmetry, we studied the impurity effect that induced by surface Rb vacancies and atomic step edges. Fig. 4**a** shows an STM image around a Rb vacancy (V$_{Rb}$) on type A surface, while Fig. 4**b** displays the tunneling spectra taken near it. The superconducting gap has an increased DOS near $E_F$ at the V$_{Rb}$ site, which evidences a local suppression of superconductivity. This gap suppression quickly disappears on moving away from V$_{Rb}$ (the Fig. 4**b** inset details the gap bottom). An exponential fit to the ZBC value as a function of distance yields a decay length of 1.41 nm (Fig. 4**c**). Similar gap suppression was also observed near the V$_{Rb}$ on type B surfaces, as shown in Figs. 4**d~f**. Since Rb vacancies are expected to be non-magnetic, this suppression of the superconductivity suggests a sign-changing pairing[55]. However, we note both d-wave and extended s-wave pairing with accidental nodes may have a sign-change, as recently suggested in the sister compound KFe$_2$As$_2$ (refs. 34, 42, 43). A further way to gain *k*-space information on the gap is to detect its response to sample boundaries. It was predicated that for a nodal pairing with a sign change, Andreev bound states at zero energy will be formed at boundaries perpendicular to the nodal direction, due to the phase change in the quasi-particles' reflection, and decay into the bulk on the scale of coherence length[56]. However, no bound states will form on boundaries perpendicular to the *anti*-nodal direction (in the ideal case). Experimentally, atomic step edges on the surface can be treated as (weak) 1D boundaries since they carry line scattering potential, and evidence of Andreev bound state has indeed been observed near [110]-oriented step edges in Bi$_2$Sr$_2$CaCu$_2$O$_{8-\delta}$ (ref. 57).

On type A surfaces we have found [110]- and [100]-oriented step edges formed during cleavage. Topographic images around these steps are shown in Fig. 5**a**. The orientations of the edges are confirmed by imaging the atomic lattice nearby (e.g. see middle panel of Fig. 5**a**), and they are all verified to be single steps with height equal to half of the c-axis lattice constant of RbFe$_2$As$_2$ (Fig. 5**b**). Figs. 5**c** and 5**e** show tunneling spectra taken along lines perpendicular to the [100] and [110] edges, respectively. To clearly see the variation of the spectra with distance to the edge, we subtracted from them a spectrum taken far away from the step edge and show the difference in Figs. 5**d** and 5**f**. The superconducting gap is suppressed in the immediate vicinity (*d*=0) of both edges, as indicated by the in-gap peaks in the difference spectra. However, on moving away from the step edge, the decay of the in-gap peaks along [110] is much slower than along the [100] direction. In Fig. 5**g**, we plot the ZBC values as function of distance from the step edge. The exponential fit yields a decay length of 6.28nm for the [110] step edge and 1.46 nm for [100]. Such a large (over 3 times) difference cannot be solely explained by the anisotropic coherence length as reflected in vortex mapping (Fig. 4), since the anisotropic ratio $\xi^L/\xi^S$ of type A surface is only 1.2 (the orientation of the vortex core relative to the step edge is indicated in Fig. 5**a**). It is more likely due to the presence of gap nodes in the {[110]} directions. As shown in Fig. 5**h**, for $d_{x2-y2}$ like pairing (left panel), a [110] boundary can give rise to bound states as it is perpendicular to the nodal direction, and these will decay into the bulk on the scale of $\xi$, while the [100] boundary cannot induce such a state. For the extended s-wave pairing suggested in refs. 34,42 (right panel), neither [110] nor [100] boundaries can induce bound states because both are perpendicular to *anti*-nodal directions. The relatively long decay length for the [110] step edge evidences the formation of bound states (although it is still shorter than the 15nm $\xi$ for this direction, as discussed below). Gap suppression is not expected for an ideal [100]-oriented edge under $d_{x2-y2}$ pairing. However, we notice that the decay length on [100] edge (1.46 nm) is very close to the decay length of gap suppression near V$_{Rb}$ (1.41 nm, see Fig.4**c**), which is measured roughly along [100] direction. Thus the gap suppression on [100] edge could be induced by random disorders near the edge (local disorders always exist on step edges, particularly for the energy-disfavored [100] edge here). We note that the point-like defect induced in-gap states are usually localized within several lattice constant to the defect site[58, 59], as theoretically their intensity varies as $1/d^2 e^{-d/\xi}$, gives a decay scale much shorter than $\xi$ (ref. 58,60).

The above observations are consistent with the $d_{x2-y2}$-like pairing suggested by several theoretical works on heavily hole-doped FeSC[16,22]. We also note some recent theoretical work has suggested that electronic nematicity is compatible with a mixing of *s + d* (or *s + id*) pairing[61-62], which may also be applicable for RbFe$_2$As$_2$. In the mixed state, signatures of *d*-wave pairing should still exist, such as the nodal gap in dI/dV and different response at the [110] and [100] edges. However, these signatures could be blurred by the s-wave component, e.g. it weakens the intensity of bound states at [110] edges and gives shorter decay length. How these theories would be modified by the (π, π) diagonal nematicity observed here requires further study.

**Effect of surface potassium (K) dosing on RbFe$_2$As$_2$:**

To gain more insights on how the (π, π) nematic state and superconductivity interplay, we hereby explore how they evolve with doping. We thus performed *in-situ* surface K dosing on RbFe$_2$As$_2$ in an STM system working at T= 4.5K (see **Methods** for details). Dosing K atoms will lower the hole-doping of the (top) FeAs layer via introducing electrons. Figs. 6 **a-c** shows typical topographic images with various K coverages ($Kc$ = 0.012-0.17 ML, and one monolayer (ML) is defined as the areal density of Fe atoms in a FeAs layer, which is ~13.4 nm$^{-2}$). Without K dosing ($Kc$ = 0), anisotropic QPI patterns that breaks (π, π)/( π, -π) equivalence were also observed at T = 4.5K, as shown in Figs. 6**f-g** (see Fig. S9 for additional data), indicating that the (π, π) nematicity *persists* above $T_c$. Upon K dosing, as shown in Fig. 6**d**, the broad DOS peak below $E_F$ is gradually suppressed as $Kc$ increases, while a gap of Δ ~3.5 meV is opened at $E_F$ at $Kc$ =0.17 ML, which is spatially uniform (Fig. 6**e**). The QPI patterns at $Kc$ =0.17 ML are shown in Figs. 6**h-i**, which remarkably became rather fourfold symmetric (see Fig. S9 for raw FFTs). This indicates that the nematicity is greatly *suppressed* in the less-hole-doped regime, compatible with the current understanding of the phase diagram. We further checked the temperature dependence of the tunneling gap at $Kc$ =0.17 ML and found it closes at T~12K (Fig. S10). Note that for $Kc$ =0.17 ML, the doping of the top FeAs layer is expected to be 0.33 holes/Fe atom (assuming each K atom can dope one electron, which could be less). Then it will be comparable to Ba$_{1-x}$K$_x$Fe$_2$As$_2$ with x≥0.66 in the hole over-doped region. Thus the observed 3.5 meV gap is most likely a superconducting gap with a $T_c$ of ~12K (which gives 2Δ/$k_B T_c$ = 6.8). Therefore, the above results directly evidenced that the (π, π) nematicity is suppressed at reduced hole doping, while the superconductivity is simultaneously enhanced.

The scattering weight in Fig. 6**i** ($Kc$ = 0.17ML) are particularly strong near the (π, 0)/(0, π) directions (dashed lines), suggesting they are from intra-band scattering of a square-shaped pocket. In Figs. 6**j** and 6**k** we plot FFT profiles taken around (π, 0) directions for $Kc$ = 0 and 0.17 ML respectively, with parabolic fitting applied. The overall dispersion for $Kc$ = 0 has a $q_F$ = 0.19 Å$^{-1}$ and $E_b$ = 25 meV, which is similar to Fig. 2**e** that measured at T=20 mK (T$_{eff}$ =310 mK) on type A surface. For $Kc$ =0.17 ML, the dispersion is also hole-like but has a significantly larger $q_F$ = 0.34 Å$^{-1}$ and $E_b$ = 50 meV (Assuming $q_F$ = 2$k_F$ for intra-band scattering, the resulting $k_F$ = 0.17 Å$^{-1}$ is comparable with that of ARPES observed α pocket of Ba$_{0.3}$K$_{0.7}$Fe$_2$As$_2$ ($k_F$~0.21 Å$^{-1}$, ref. 36)). However, a reduced $q_F$ at $Kc$ = 0 is apparently unexpected, as it has a higher hole concentration than the K-dosed case. It would then imply certain band reconstructions happened in the presence of (π, π) nematicity at $Kc$ = 0.

**Discussions:**

Our measurement evidences a novel (π, π) diagonal nematicity that coexists with a nodal $d_{x2-y2}$ like pairing component in the strongly hole-doped RbFe$_2$As$_2$ (with a 3d$^{5.5}$ configuration). The diagonal nematicity persists into the normal state above Tc, which implies that the superconductivity emerges in the nematic phase, and nematic fluctuations may play a role. Furthermore, the surface electron doping clearly demonstrated an anti-correlation between such a nematicity and superconductivity. In Fig. 7 we summarize above results in a global phase diagram of 122 type iron-pnictides. These results give important clues on the origin of nematicity, and highlight the intimate relation between nematicity and superconducting pairing. As shown in Fig. 7, the stripe AFM and (π, 0) nematicity of the parent material weakened through hole-doping, and disappear near the "optimal" doping with highest $T_c$. The pairing at optimal doping is widely believed to be $s_\pm$ wave and mediated by (π, 0) spin fluctuations. Here in the over-hole-doped regime, the re-emergence of a new type of diagonal nematic state and change in the pairing symmetry with suppressed $T_c$ imply the system approaches a new regime with different fluctuations. Since an AFM Mott insulating state resembling the parent compound of cuprates was predicted for $3d^5$ configuration, the related spin/nematic fluctuations are likely the candidate. It may give rise to or mediate both the diagonal nematicity and the d-wave pairing component, although the microscopic details require further theoretical refinement. Therefore, our results would lay the groundwork for a unified understanding of the cuprates and FeSCs.

While preparing this paper, we became aware of an NMR study on $CsFe_2As_2$ which also suggests a nematic state along the (π, π) direction[63]. It thus provided complementary evidences for the re-emergence of electronic nematicity in heavily hole-doped FeSCs. We also noticed that another NMR study[64] has revealed strong spin fluctuations in $RbFe_2As_2$ and $CsFe_2As_2$.

**Methods:**

The $RbFe_2As_2$ singles crystals were grown in alumina crucibles by a self-flux method with $T_c$ ~2.5K (ref. 27) (see Fig. S1). STM experiment (except the surface K dosing effect) was conducted in a commercial $^3He/^4He$ dilution refrigerator STM (Unisoku) at the base temperature of ~20 mK. The effective electron temperature ($T_{eff}$) of this system was checked to be ≤310 mK by measuring the superconducting gap of Al films (see Fig. S2). $RbFe_2As_2$ samples were cleaved in ultrahigh vacuum at ~80 K (liquid nitrogen temperature) and immediately transferred to the STM module. Pt tips were used after careful treatment on Au (111) sample. The tunneling spectroscopy (dI/dV) was performed using a standard lock-in technique with modulation frequency $f$ = 787 Hz, and the modulation amplitudes (ΔV) are specified in the figure captions.

The surface K dosing on $RbFe_2As_2$ was conducted in another cryogenic STM at 4.5K. $RbFe_2As_2$ sample was cleaved at ~30K and Pt tips were also used after treatment on Au (111). K atoms were evaporated from standard SAES alkali metal dispensers, and the sample was kept at 80K during deposition. The deposition rate was carefully calibrated by directly counting surface K atoms at low coverage. The tunneling spectra were obtained by using lock-in technique with modulation frequency $f$ = 915 Hz and amplitude ΔV = 1 mV.

**Acknowledgments:**

We thank Professor P. Adelmann for the help on sample growth, and thank Professors T. Wu, J. P. Hu, D. H. Lee and Dr. D. Peets for helpful discussions. This work is supported by the National Natural Science Foundation of China, National Key R&D Program of the MOST of China (Grant No. 2016YFA0300200 and 2017YFA0303004), National Basic Research Program of China (973 Program) under grant No. 2015CB921700, and Science Challenge Project (grant no. TZ2016004).

**Author contributions**

The growth of $RbFe_2As_2$ singles crystals was performed by T. Wolf. The STM measurements and data analysis were performed by X. Liu, R. Tao, M. Ren, W. Chen and T. Zhang. T. Zhang and D. L. Feng coordinated the project and wrote the manuscript. All authors have discussed the results and the interpretation.

## Additional information

Competing financial interests: The authors declare no competing financial interests.

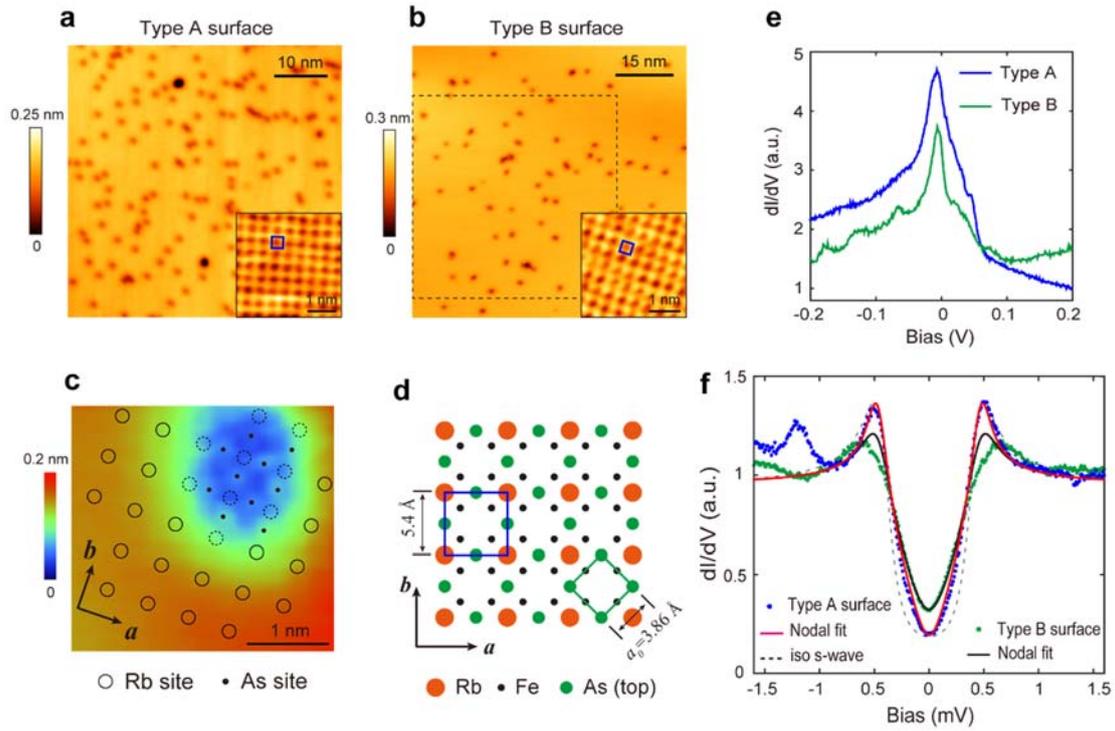

**Figure 1 | Surface atomic structure and tunneling spectrum of cleaved RbFe$_2$As$_2$.** (**a**) Topographic image of type A surface (55×55nm$^2$, $V_b$ = 150mV, I = 100pA), inset is atomically resolved image of a defect-free area, showing a square lattice with a lattice constant of 5.4 Å. (**b**) Topographic image of type B surface (75×75nm$^2$, $V_b$ = 1V, I = 10pA). The inset is an atomically resolved image of a defect-free area, showing a similar lattice to type A surface. QPI mapping on type B surface was performed in the dashed square. (**c**) A closer image of a large basin defect ($V_b$ = 6mV, I = 3nA) reveals a different atomic lattice inside the basin. We attribute these atoms to the As layer beneath the surface Rb layer. (**d**) Sketch of surface atomic structure. The surface is half covered by Rb and forms a √2×√2 (R45°) lattice with respect to the top As layer. (**e**) dI/dV spectra on both surfaces (set point: type A surface: $V_b$=200mV, I=100pA, ΔV = 1mV; type B surface: $V_b$ =200mV, I=50pA, ΔV = 1mV). A DOS peak is observed slightly below $E_F$. (**f**) Low-energy dI/dV spectra taken on type A and B surfaces ($V_b$ =2mV, I=100pA, ΔV=30μV, $T_{eff}$ = 310 mK), and fits to nodal gap and isotropic s-wave gap functions. All the data shown in this figure are taken at T = 20 mK ($T_{eff}$ = 310 mK).

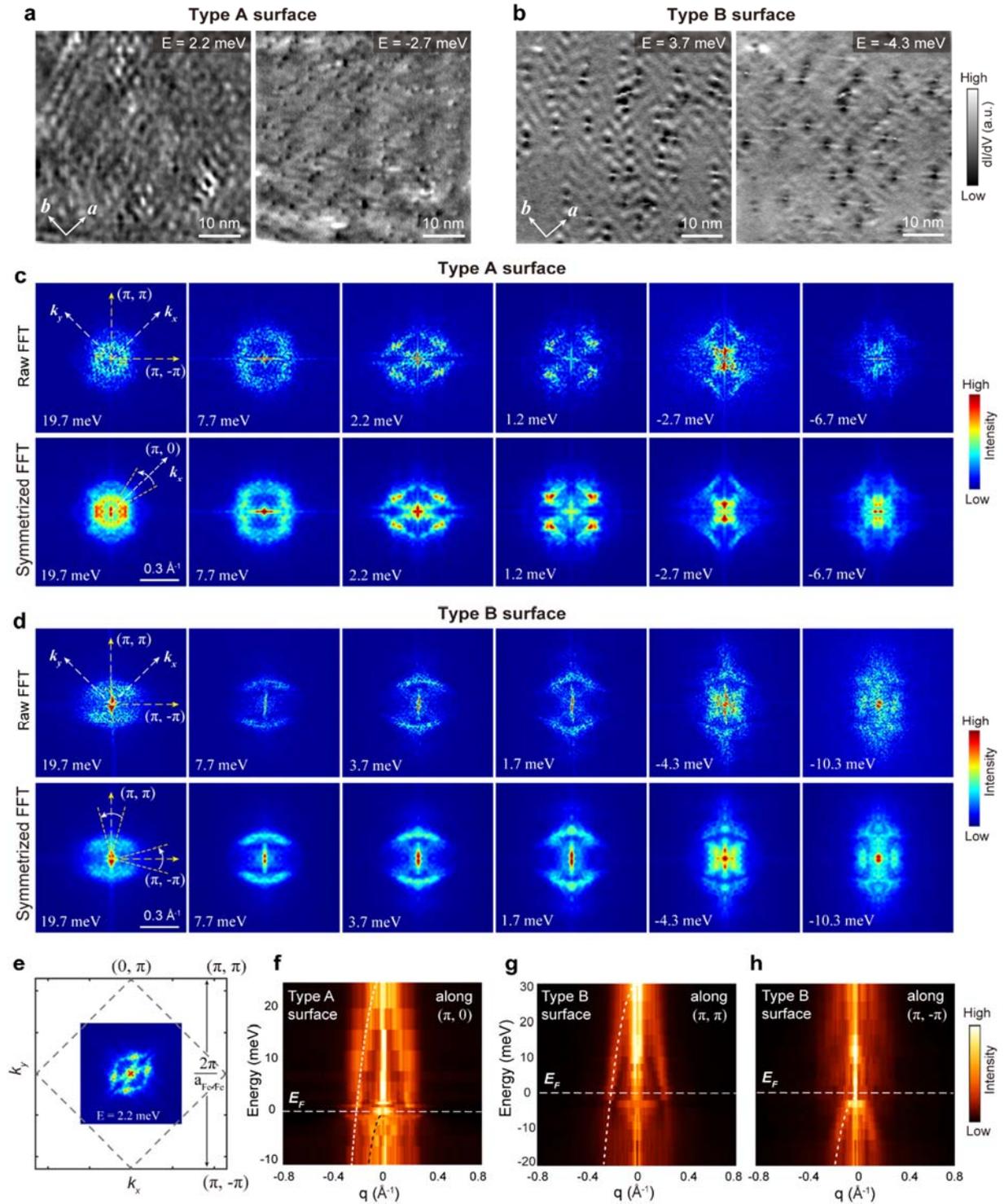

**Figure 2 | QPI measurements on RbFe$_2$As$_2$.** (**a**, **b**) Representative dI/dV maps taken on type A and type B surfaces, respectively. The mapping area of panel **a** is the same as shown Fig. 1**a**. The mapping area of panel **b** is marked in Fig. 1**b**. (**c, d**) Representative raw and symmetrized FFT images of the dI/dV maps taken on type A and type B surfaces, respectively. The orientation of the unfolded BZ is marked on the first image of each panel. (**e**) Sketch of the unfolded Brillouin zone of RbFe$_2$As$_2$ and its relation to the FFT pattern. The C$_4$ symmetry breaking makes (π, π) and (π, -π) inequivalent. (**f**) FFT profiles along (π, 0) direction of type A surface (averaged over a 30° angle that indicated in panel **c**), in which a hole-like dispersion can be observed. Parabolic fit (white dashed curve) gives $E_b$ = 27 meV and $q_F$ = 0.21 Å$^{-1}$. Another hole-like dispersion (black dashed curve) slightly below $E_F$ is observed. (**g**) FFT profile of type B surface along the (π, π) direction (averaged over a 30° angle indicated in panel **d**). (**h**) FFT profile of type B surface along the (π, -π) direction

(averaged over a 30° angle), another hole-like band can be seen below $E_F$. (Note: All QPI data are taken at T = 20 mK ($T_{eff}$ = 310 mK). The symmetrized FFTs are mirror symmetrized along (π, π) and (π, -π) directions. Each dI/dV maps are taken at a $V_b$ equal to the mapping energy (labeled) and I = 100pA; lock-in modulation (ΔV) for each map has an amplitude of 5% $V_b$).

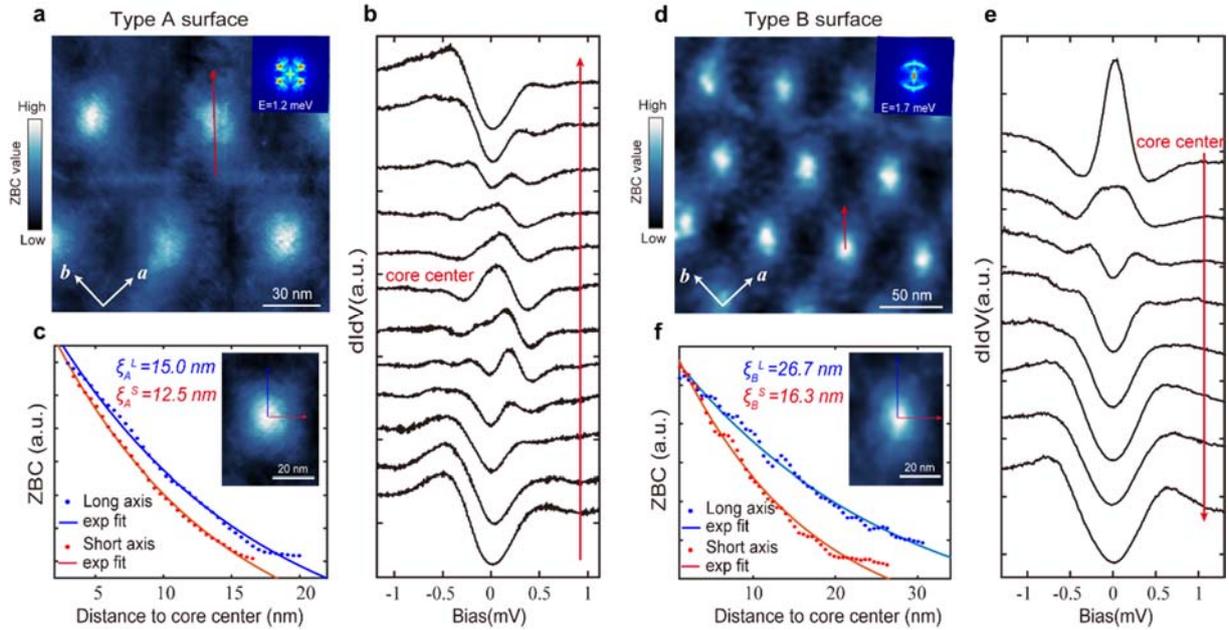

**Figure 3 | Magnetic vortex mapping on RbFe$_2$As$_2$.** (**a**) Zero bias conductance (ZBC) mapping on type A surface under *B* = 0.5T (size: 150×150 nm$^2$; $V_b$ = 1.2mV, I = 150pA, ΔV = 50μV). A spatially averaged core is shown in panel **c**. Inset is an FFT image aligned to the ZBC map. (**b**) Evolution of the dI/dV spectra taken across the vortex core ($V_b$ = 1.5mV, I = 100pA, ΔV = 50μV), along the red arrow in **a**. A zero bias peak is observed at the center and splits when leaving the core. (**c**) ZBC line profiles along the long and short axes of an averaged vortex core (inset). Solid curves are exponential fits which yield the coherence lengths $\xi^L$ and $\xi^S$. (**d**) ZBC mapping of type B surface under *B* = 0.5T (size: 225×225 nm$^2$; $V_b$ = 1.5mV, I = 120pA, ΔV = 50μV). The elongated direction of the vortex cores is where the QPI shows arc-like features (inset). (**e**) dI/dV spectra taken across the vortex core ($V_b$ = 1.5mV, I = 100pA, ΔV = 50μV), along the red arrow in **d** – a zero bias peak is also observed. (**f**) ZBC line profiles along the long and short axes of a vortex core on type B surface (inset) and their exponential fits. The difference between $\xi^L$ and $\xi^S$ on type B surface is more significant than on type A surface. (All the data shown in this figure are taken at T = 20 mK ($T_{eff}$ = 310mK)).

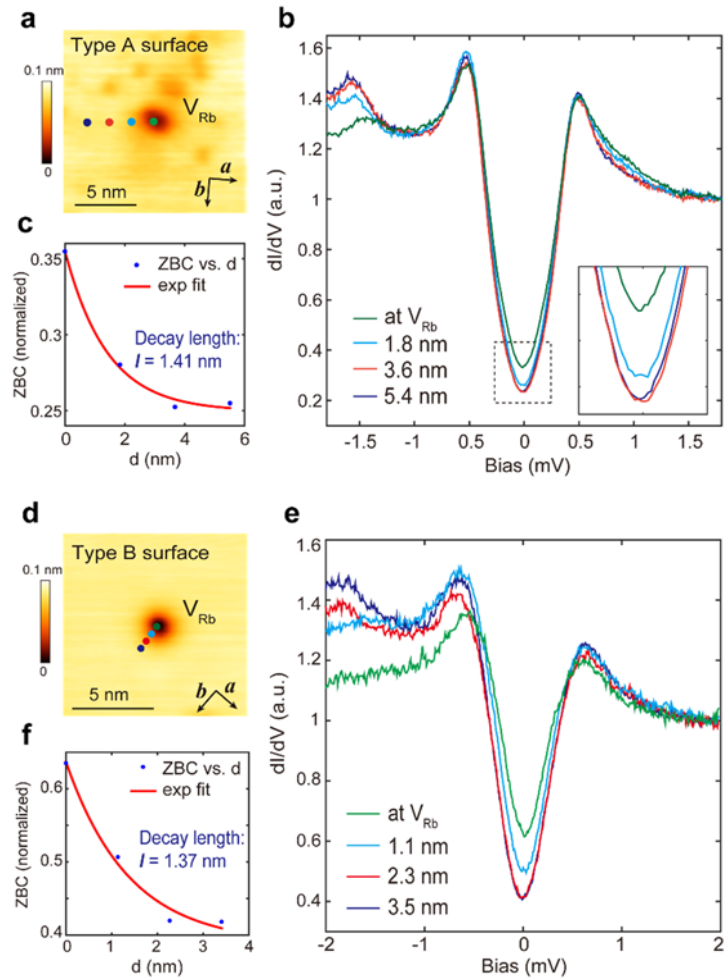

**Figure 4 | Effect of a surface defect ($V_{Rb}$) on superconductivity** (**a**) Topographic image of a surface Rb vacancy ($V_{Rb}$) on type A surface ($V_b$ = 30mV, I = 10pA). (**b**) dI/dV spectra taken at various distances from $V_{Rb}$ ($V_b$ = 1.8mV, I = 100pA, $\Delta V$ = 50µV, $T_{eff}$ = 310mK), the tip positions are marked in panel **a**. Inset displays the gap bottom showing the suppression of the gap. (**c**) ZBC value of the gap bottom as function of distance from $V_{Rb}$; the red curve is an exponential fit which yields a decay length of 1.41 nm. (**d**) Topographic image of a Rb vacancy on type B surface ($V_b$ = 1.0V, I = 10pA). (**e**) dI/dV spectra taken at various distances from $V_{Rb}$ ($V_b$ = 2mV, I = 100pA, $\Delta V$ = 50µV, $T_{eff}$ = 310mK), the tip positions are marked in panel **d**. (**f**) ZBC value as function of distance from $V_{Rb}$; red curve is an exponential fit which yields a decay length of 1.37nm.

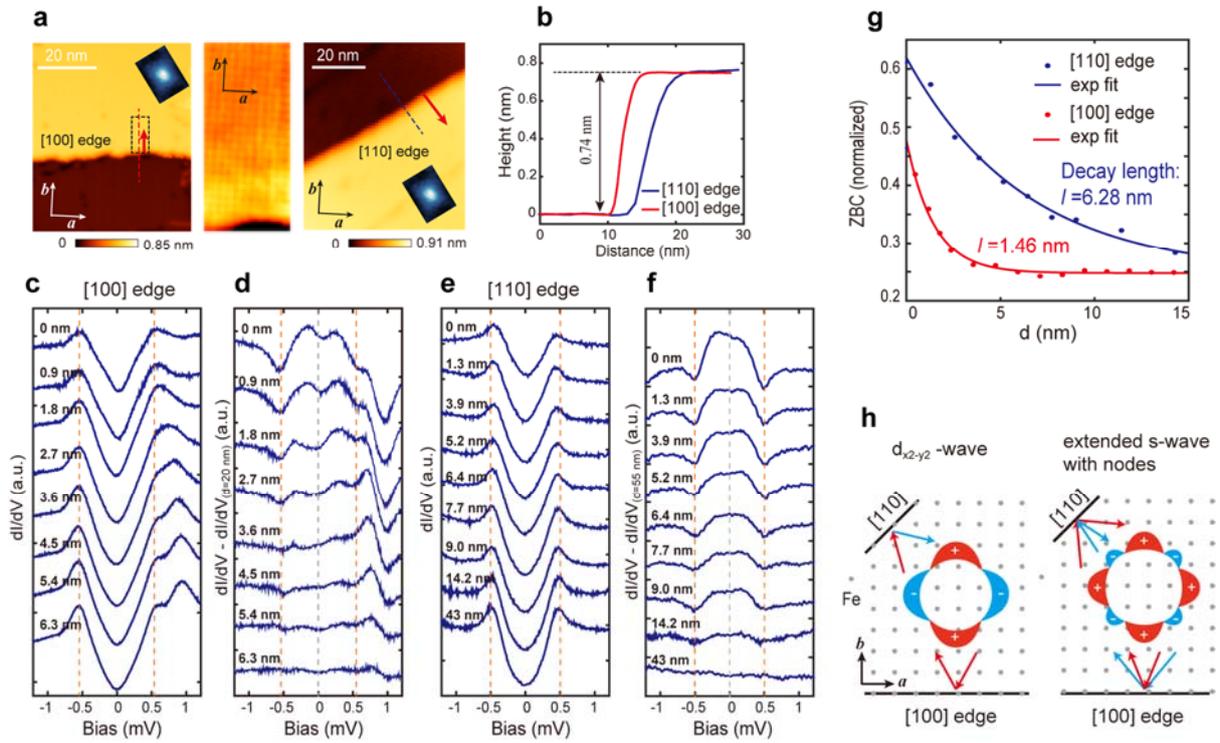

**Figure 5 | Effect of atomic step edges on superconductivity.** (**a**) Topography of the step edges along the [110] and [100] directions, as marked in the image (left and right panel: $V_b$ = 0.5V, I = 10pA, middle panel: $V_b$ = 2mV, I = 200pA). Their orientation relative to the $C_4$ symmetry breaking is indicated by the inserted vortex core image. The middle panel is an atomically resolved image near the step edge (taken in dash rectangle in **a**). (**b**) Line profiles along the dashed lines marked in panel **a**, showing that both steps are half a unit cell high. (**c-d**) Superconducting gap evolution when leaving the [100] step edge (along red arrow in the first panel of **a**, setpoint: $V_b$ = 1.5mV, I = 80pA, ΔV = 50μV). For clarity, a gap far from the step edge has been subtracted for panel **d**. Dashed lines mark the position of coherence peaks and $E_F$. (**e-f**) Superconducting gap evolution when leaving the [110] step edge (along red arrow in the third panel of **a**, setpoint: $V_b$ = 1.5mV, I = 100pA, ΔV = 50μV) and the subtracted spectra. (**g**) ZBC decay evolution when leaving different steps (dots) and their exponential fits (solid lines). (**h**) Demonstration of the quasi-particle scattering on step edges with different orientation. Left panel: $d_{x^2-y^2}$-wave pairing. Right panel: extended-s wave pairing with eight accidental nodes. (All the data shown in this figure are taken at T = 20 mK ($T_{eff}$ = 310 mK)).

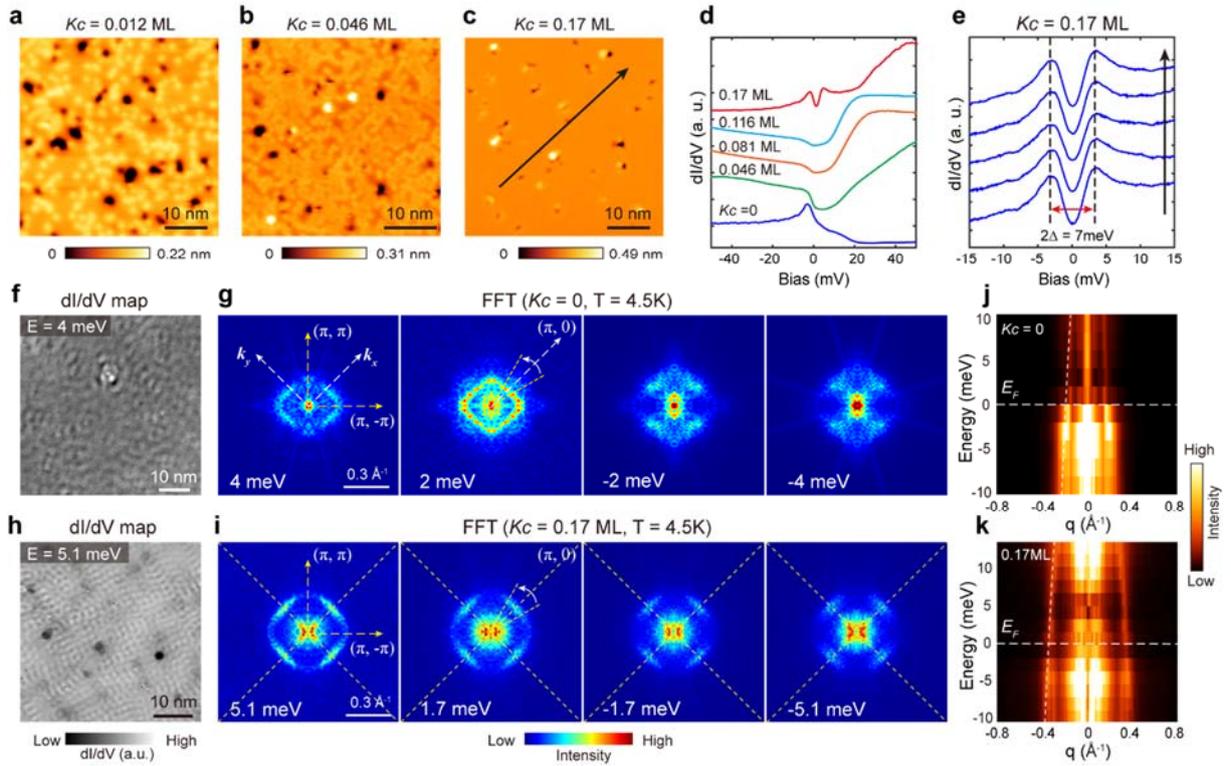

**Figure 6 | Effect of surface potassium (K) dosing on RbFe$_2$As$_2$.** (**a-c**) Topographic images of RbFe$_2$As$_2$ with various K coverage ($Kc$ = 0.012ML, 0.046ML, 0.17ML, respectively). K atoms appear as random bright spots at low $Kc$. (Setpoints of **a**: $V_b$ =1V, I = 50pA; **b**: $V_b$ =1V, I = 20pA; **c**: $V_b$ =0.2V, I = 200pA). (**d**) Spatially averaged dI/dV spectra of RbFe$_2$As$_2$ with various $Kc$ ($V_b$ = 50mV, I = 200pA, $\Delta$V = 1mV, T = 4.5K for all the spectra). (**e**) Low energy dI/dV spectra of RbFe$_2$As$_2$ at $Kc$ = 0.17ML ($V_b$ = 15mV, I = 300pA, $\Delta$V = 1mV, T = 4.5K), taken along the arrow in panel **c**. A spatially uniform gap with the size of 2$\Delta$ = 7.0 meV is observed. (**f**) A representative dI/dV map taken at $Kc$ = 0, T = 4.5K ($V_b$ = 10mV, I = 200pA, $\Delta$V = 1mV) (**g**) Selected FFT images of dI/dV maps with $Kc$ = 0. The FFTs are mirror symmetrized along ($\pi$, $\pi$) and ($\pi$, -$\pi$) directions. (**h**) A representative dI/dV map taken at $Kc$ = 0.17ML ($V_b$ = 10mV, I = 200pA, $\Delta$V = 1mV, T = 4.5K) (**i**) Selected FFT images of the dI/dV maps with $Kc$ = 0.17ML; FFTs are also mirror symmetrized along ($\pi$, $\pi$) and ($\pi$, -$\pi$) directions. (**j**) FFT profiles along the ($\pi$, 0) direction for $Kc$ = 0 (averaged over a 30° angle, as indicated in the second panel of **g**). Parabolic fit (dashed curve) gives $E_b$ = 25 meV and $q_F$ = 0.19 Å$^{-1}$. (**k**) FFT profiles along the ($\pi$, 0) direction (averaged over a 30° angle) for $Kc$ = 0.17ML. Parabolic fit (dashed curve) gives $E_b$ = 50 meV and $q_F$ = 0.34 Å$^{-1}$.

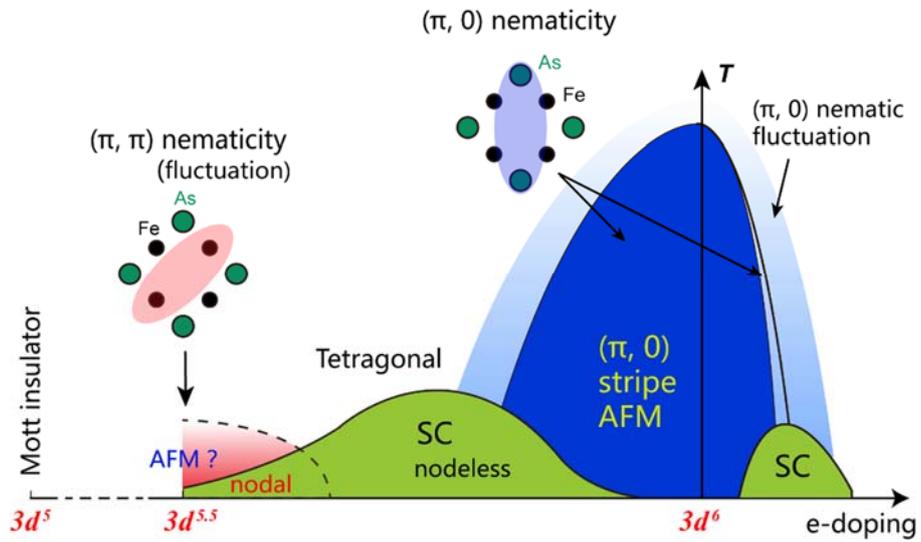

**Figure 7 | Schematic phase diagram of 122 type iron-pnictides.** The blue area denotes (π, 0) stripe AFM which coexists with the (π, 0) nematicity, and the light blue area denotes (π, 0) nematic fluctuation. The two green areas are superconducting dome. The red area denotes (π, π) nematicity at the $3d^{5.5}$ configuration (which may become (π, π) nematic fluctuation at increased temperature). For the hole-doping superconducting dome, electron pairing is nodeless (s±) in the middle of the dome, but became nodal ($d_{x^2-y^2}$ like) as approaching $3d^{5.5}$. A Mott insulator phase is theoretically expected at the $3d^5$ configuration. The upper insets demonstrate the symmetry breaking in (π, 0) and (π, π) nematic state.

# Supplementary Materials for

# Evidence of nematic order and nodal superconducting gap along [110] direction in RbFe$_2$As$_2$

Xi Liu[1+], Ran Tao[1+], Mingqiang Ren[1+], Wei Chen[1], Qi Yao[1], Thomas Wolf[3], Yajun Yan[1], Tong Zhang[1,2*], Donglai Feng[1,2*]

[1]State Key Laboratory of Surface Physics, Department of Physics, and Advanced Materials Laboratory, Fudan University, Shanghai, 200433, China

[2]Collaborative Innovation Center of Advanced Microstructures, Nanjing, 210093, China

[3]Institute for Solid State Physics, Karlsruhe Institute of Technology, D-76021 Karlsruhe, Germany

+ These authors contributed equally to this work.
*Email: tzhang18@fudan.edu.cn, dlfeng@fudan.edu.cn

## I. Transport measurements of RbFe$_2$As$_2$ single crystal

RbFe$_2$As$_2$ single crystals with the ThCr$_2$Si$_2$-type structure (Fig.S1**a**) were grown in alumina crucibles by a self-flux method (as described in ref. 27 of the main text). Figs. S1**b-c** shows the temperature dependence of the sample resistance. Zero resistance is observed below $T_c$ ~2.5K (Fig. S1**c**). Above $T_c$ the resistance obeys a $T^2$ power law up to 50K. By fitting the data to $R(T) = R_0 + kT^2$ for T<50K (red curve in Fig. S1**b**), we obtain the residual resistivity ratio, RRR= $R(300 K)/R_0$ = 533. Such a large RRR value has been widely reported for AFe$_2$As$_2$ (A=K, Rb, Cs), which was attributed to a strong correlation induced coherence-incoherence crossover at high temperatures (ref. 26 in main text).

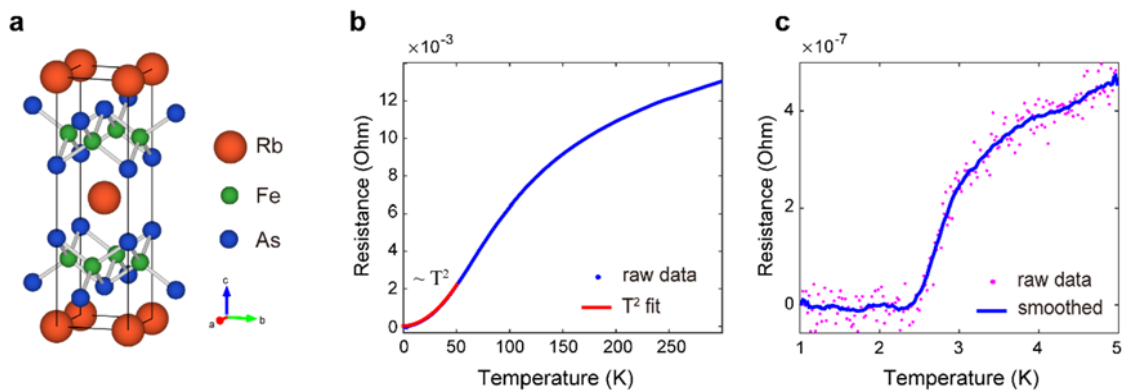

**Figure S1** | (**a**) Bulk crystal structure of RbFe$_2$As$_2$. (**b**) Temperature dependence of the resistance (0~300K), the red curve below 50 K is a fit to $T^2$. (**c**) Temperature dependence of the resistance across the superconducting transition at $T_c$ ~2.5K.

## II. Calibration of the effective electron temperature of the STM system

Due to the electrical noise and RF radiations, the effective electron temperature ($T_{eff}$) of a low-$T$ STM is usually higher than the thermometer reading. The $T_{eff}$ of the dilution refrigerator STM used in this work is calibrated by measuring the superconducting gap of an Al film grown Si (111). Fig. S2**a** shows a typical STM image of the Al/Si(111) film, with a thickness of ~ 20 monolayer (ML), and Fig. S2**b** shows its superconducting gap spectrum. A standard BCS fit (red curve) yields $\Delta$ = 0.19 meV and $T_{eff}$ = 310 mK. Here we note that in order to make a conservative estimation of $T_{eff}$, Dynes broadening term is not used in the fitting ($\Gamma$ = 0). In the presence of finite $\Gamma$ or other broadening factors, the actual $T_{eff}$ could be slightly lower than the fitted value.

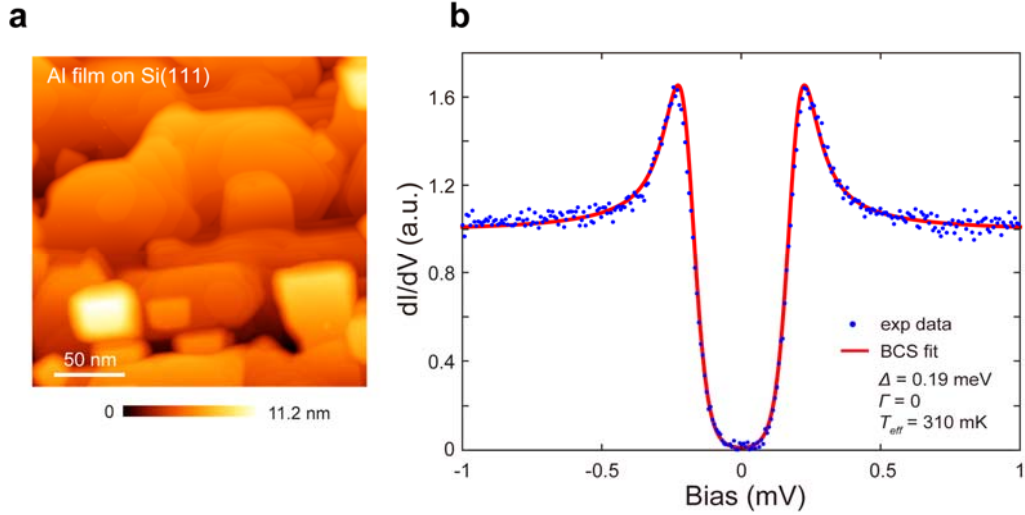

**Figure S2 |** (**a**) Topographic image of an Al/Si(111) film of thickness ~20ML. (**b**) The superconducting gap of the Al/Si(111) film taken at T = 20mK ($V_b$ = 1mV, I = 100pA, $\Delta V$ = 30µV). Red curve is the BCS fit with $\Delta$ = 0.19 meV, $T_{eff}$ = 310 mK and $\Gamma$ = 0.

## III. Determining the surface atomic structure of cleaved RbFe$_2$As$_2$

The surface structure of cleaved RbFe$_2$As$_2$ is revealed by resolving the atomic lattice of the defect-free area and that inside of the Rb vacancies. Fig. S3**a** is a typical image of type B surface with multiple Rb vacancies. Fig. S3**b** shows the region marked in Fig. S3**a** in greater detail (shown with a linear color scale). A square lattice inside of the Rb vacancies can be seen and has a lattice constant of ~3.8 Å, matching $a_0$, which is mostly likely from the underlying FeAs layer. The lattice of the defect-free area is hard to see in Fig. S3**b** due to its much smaller corrugation. To enhance the contrast, a non-linear color mapping is used in Fig. S3**c**, in which both the surface Rb lattice and lattice inside the vacancy can be seen. By comparing the atomic sites of these two lattices (as marked in Fig. S3**d**), the surface lattice model is derived and shown in Fig. S3**e**. The surface Rb forms a √2×√2 (R45°) lattice with respect to the As lattice.

For the lattice model in Fig. S3**e**, the surface Rb atoms will have another set of occupation sites, as illustrated by the dashed circles. These two equivalent occupation sites are shifted by 1/2 unit cell with respect to each other (along both **a** and **b** directions), which should result in domain structures when both are present. We indeed observed such domain structures on samples cleaved at a lower temperature (~30K), as shown in Fig. S4**a**. There are domain boundaries running through the surface (marked by red arrows). Fig. S4**b** is an atomically resolved image near a domain boundary. One can see aside of the boundary, the surface still displays a √2×√2 lattice. However, the lattice of the upper domain is shifted by 1/2 unit cell along **a** and **b** with respect to the lower domain. To illustrate this, we draw a lattice of white spots which matches the atomic lattice of the lower domain, however it mismatches the upper domain by the above offset. The existence of different domains gives further support to the surface lattice model in Fig. S3**e**.

The assignment of surface atomic structure above is based on STM imaging. To further confirm the orientation of surface √2×√2 lattice with respect to the bulk FeAs lattice, we performed Laue diffraction measurement to accompany STM imaging. The results are summarized in Fig. S5. We first determined the orientation of $a_0$ and $b_0$ (the in-plane basic vectors of **2Fe** unit cell) of RbFe$_2$As$_2$ single crystal by comparing its measured Laue pattern with a simulated pattern. As shown in the simulation in Fig. S5**b**, the four diffraction spots that closest to the center, labelled by red colored Miller indices, are originated from {10X} plane family (defined by **2Fe** unit cell, so they are along the $a_0$ or $b_0$ directions); while the spots with blue colored Miller indices are from {11X} plane family. In the measured Laue pattern, only the spots close to the center show up with significant weight (Fig. S5**a**), so the crystal orientation can be determined by comparing it to Fig. S5**b**. Then the crystal was glued on STM sample holder with $a_0$ and $b_0$ aligned to X and Y scan directions, respectively. The STM image of cleaved surface (Fig. S5**c**) then directly shows the surface √2×√2 lattice is rotated 45° with respect to $a_0$ and $b_0$. We also repeated the same procedure on a pure FeSe single crystal, the results are shown in Figs. S5**d-f**. The FeSe has a Laue pattern in analogous to RbFe$_2$As$_2$, and its $a_0$, $b_0$ directions are determined in a similar way. The STM image in Fig. S5**f** show that its surface lattice has a constant of 3.75 Å, with the direction the same as $a_0$ and $b_0$. This is well expected for a Se terminated surface of FeSe. Therefore, combined Laue and STM measurement directly indicate the surface √2×√2 lattice of RbFe$_2$As$_2$ is rotated 45° with respect to $a_0$, $b_0$.

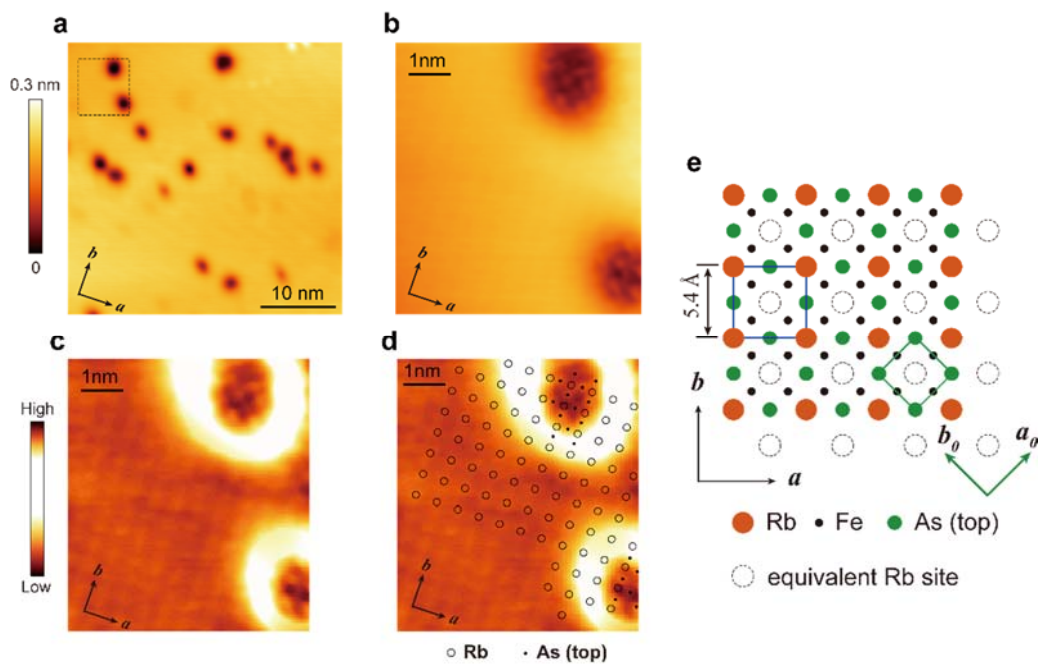

**Figure S3** | (**a**) An STM image of type B surface ($V_b$ = 0.8V, I = 100pA) B. (**b**) Higher-resolution image ($V_b$ = 6mV, I = 3nA) of the region marked in panel **a**, shown with a linear color scale. (**c**) The same image as panel **b** but mapped with a non-linear color scale to reveal the Rb lattice. (**d**) The same image as panel **c**, with the atomic sites marked. (**e**) The surface lattice model derived from panel **d**. The orientation of 2Fe unit cell is denoted by $a_0$, $b_0$.

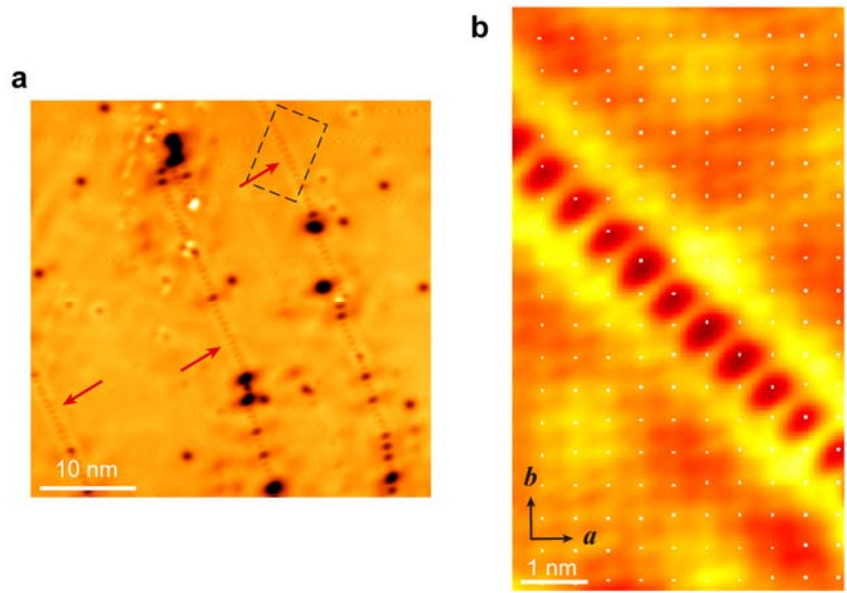

**Figure S4 |** (**a**) STM image of RbFe$_2$As$_2$ surface cleaved at T~30K (V$_b$ = -4meV, I = 100pA). Domain boundaries are indicated by red arrows. (**b**) Atomically resolved STM image (V$_b$ = -4meV, I = 500 pA) taken in the region marked in panel **a**. A lattice of white spots is intentionally drawn to match the atomic lattice of the lower domain. However it is misaligned from the upper domain lattice by 1/2 unit cell along **a** and **b**. (Images are taken at T=4.5K)

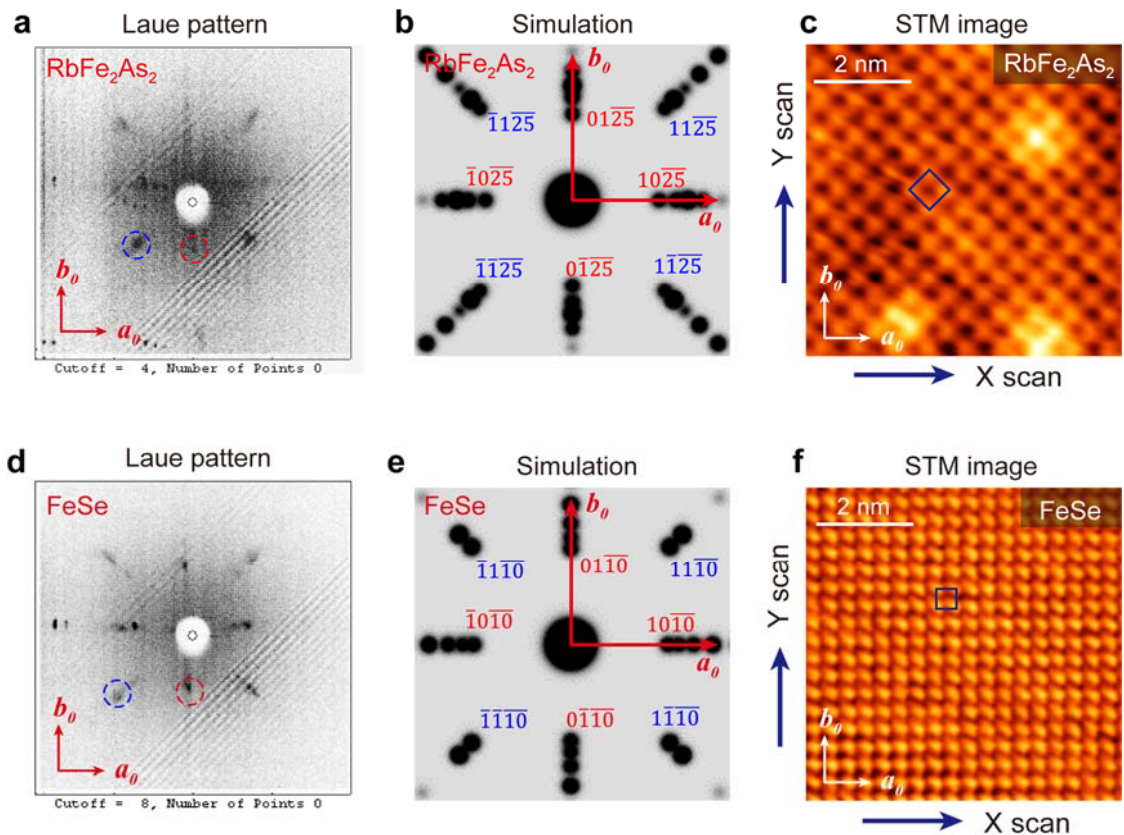

**Figure S5 | Determine the surface lattice orientation of RbFe$_2$As$_2$ by Laue diffraction and STM imaging.** (**a**) Laue diffraction pattern of RbFe$_2$As$_2$ single crystal. (**b**) Simulated Laue pattern using SingleCrystal$^{TM}$ software for RbFe$_2$As$_2$ with **a$_0$**, **b$_0$** along X and Y directions. The Miller indices (hkl) of the planes corresponding to the diffraction points near center are marked, which are belong to {01X} (red) and

{11X} (blue) plane families. (**c**) STM image of cleaved RbFe$_2$As$_2$ with $a_0$, $b_0$ pre-aligned to X scan and Y scan directions ($V_b$ = 8mV, I = 1nA). The lattice constant is measured to be 5.4 Å, and the lattice is rotated 45° with respect to $a_0$, $b_0$. (**d**) Laue diffraction pattern of a FeSe single crystal. (**e**) Simulated Laue pattern of FeSe with $a_0$, $b_0$ along X and Y directions. The Miller indices (hkl) of the plane corresponding to the diffraction points near center are also marked. (**f**) STM image of cleaved FeSe crystal with $a_0$, $b_0$ pre-aligned to X scan and Y scan directions ($V_b$ = 100mV, I = 100pA). The lattice constant is measured to be 3.75 Å, and the lattice has the same orientation with $a_0$, $b_0$. Note: The spots marked by red and blue circles in panels **a**, **d** are corresponding to the spots marked by the same colored indices in the simulated pattern (in panels **b**, **e**). They originated from {10X} and {11X} plane families, respectively.

## IV. Spatial dependence of the superconducting gap

The superconducting gap of RbFe$_2$As$_2$ is checked to be spatially homogenous, as shown in Fig. S6 below.

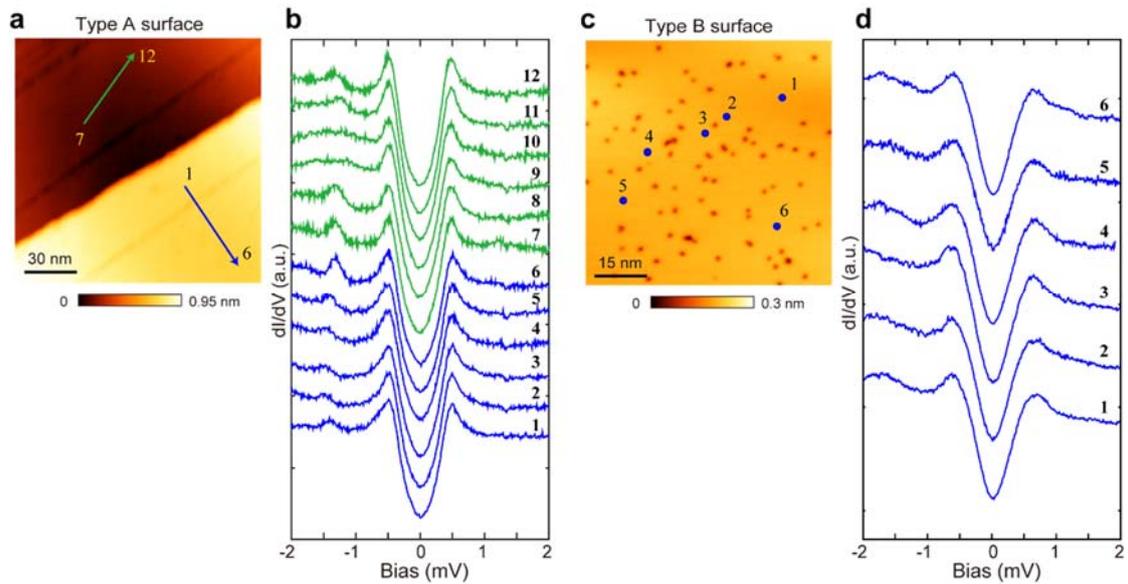

**Figure S6 | Spatial dependence of the superconducting gap on type A and B surfaces.** (**a**) STM image of a type A surface ($V_b$ = 0.5V, I=10 pA) (**b**) superconducting gap spectra taken along the arrows marked in panel **a** ($V_b$ = 2mV, I = 100pA, ΔV = 50μV). (**c**) STM image of type B surface ($V_b$ = 1V, I = 10pA). (**d**) Superconducting gap spectra taken at the spots marked in panel **c** ($V_b$ = 2mV, I = 100pA, ΔV = 50μV). The spots are randomly chosen. All the spectra shown in this figure are taken at T = 20mK.

## V. Additional data of Quasi-particle interference (QPI)

Additional dI/dV maps, raw FFTs and symmetrized FFTs taken on type A and B surfaces are shown in Fig. S7 and Fig. S8, respectively. Symmetrized FFTs are all obtained by mirror symmetrizing the raw FFTs along (π, π) and (π, -π) directions. Detailed process is 1): Identify the (π, π) and (π, -π) directions of the raw FFT by using atomically resolved images; 2): Mirror flip the raw FFT along (π, π) and add it to the raw FFT; 3) Flip the results of 2) along (π, -π) and add it to the results of 2).

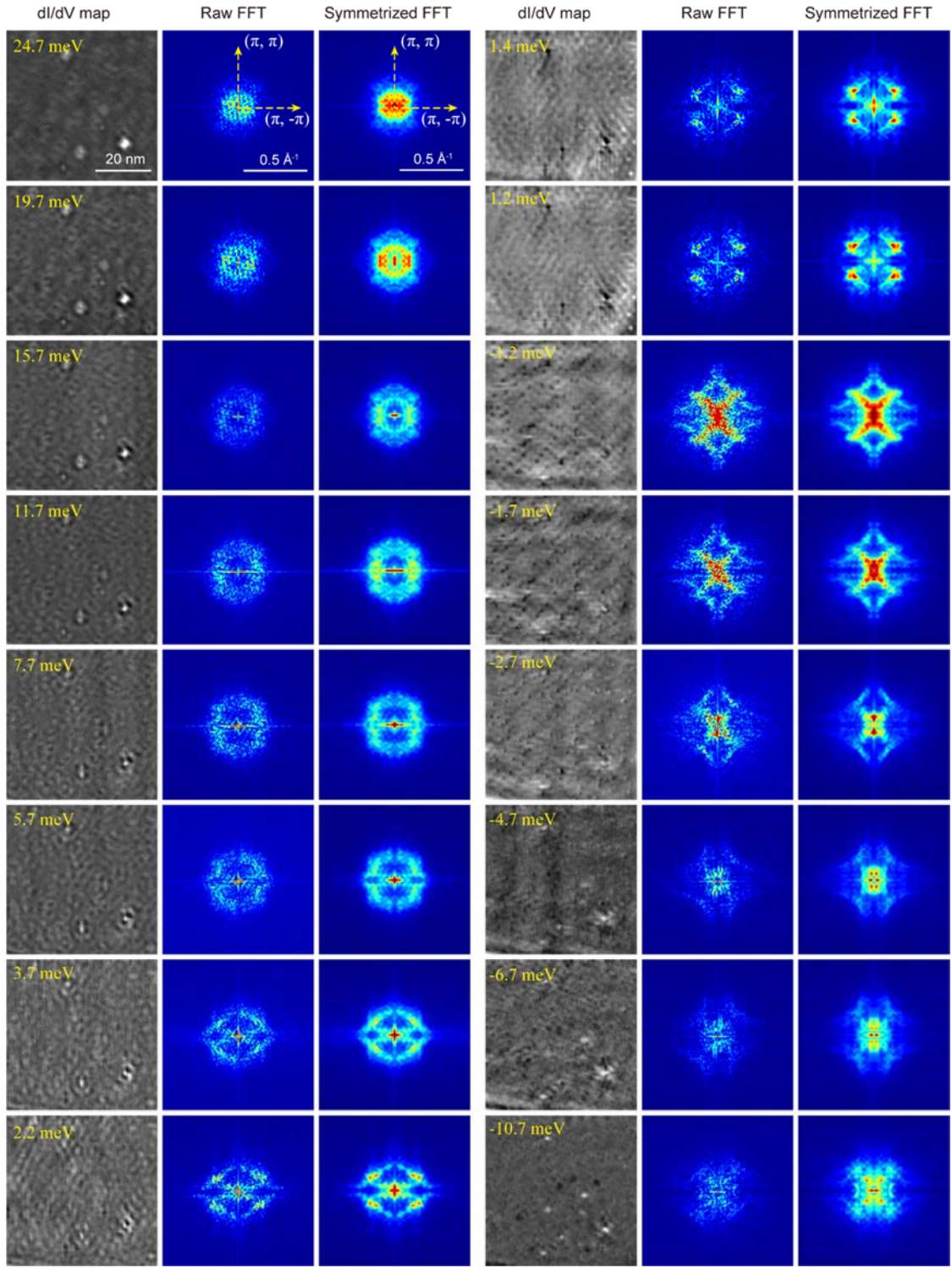

**Figure S7 | A full set of dI/dV maps taken on type A surface and their FFTs.** The topography of the mapping area is shown in Fig. 1**b**. Each dI/dV maps are taken at a $V_b$ equal to the mapping energy (labeled on the map) and I = 100pA; the lock-in modulation (ΔV) for each map has an amplitude of 5% $V_b$. Details about the FFT symmetrization is described above. Every dI/dV map has 256 × 256 pixels.

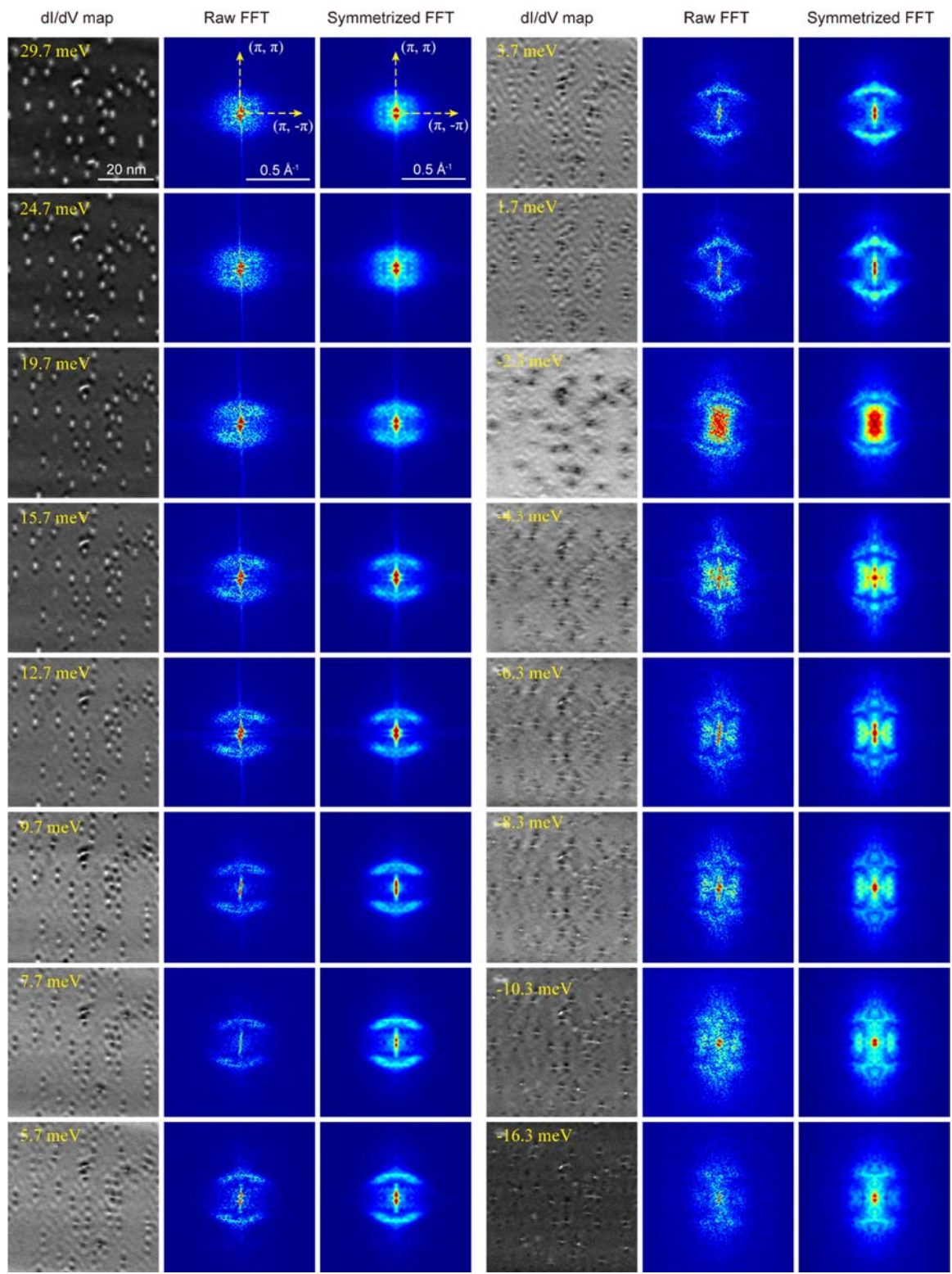

**Figure S8 | A full set of dI/dV maps taken on type B surface and their FFTs.** The topography of the mapping area is shown in Fig. **1c**. Each dI/dV maps are taken at a $V_b$ equal to the mapping energy (labeled on the map) and I = 100pA; the lock-in modulation (ΔV) for each map has an amplitude of 5% $V_b$. Every map has 256 × 256 pixels.

## VI. Additional data of surface K-dosed RbFe$_2$As$_2$

Additional dI/dV maps, raw FFTs and symmetrized FFTs taken on RbFe$_2$As$_2$ with Kc =0 and Kc =0.17 ML at T= 4.5K are shown in Fig. S10. The symmetrized FFTs are also obtained by mirror symmetrizing the raw FFTs along (π, π) and (π, -π) directions, as described in section **V**.

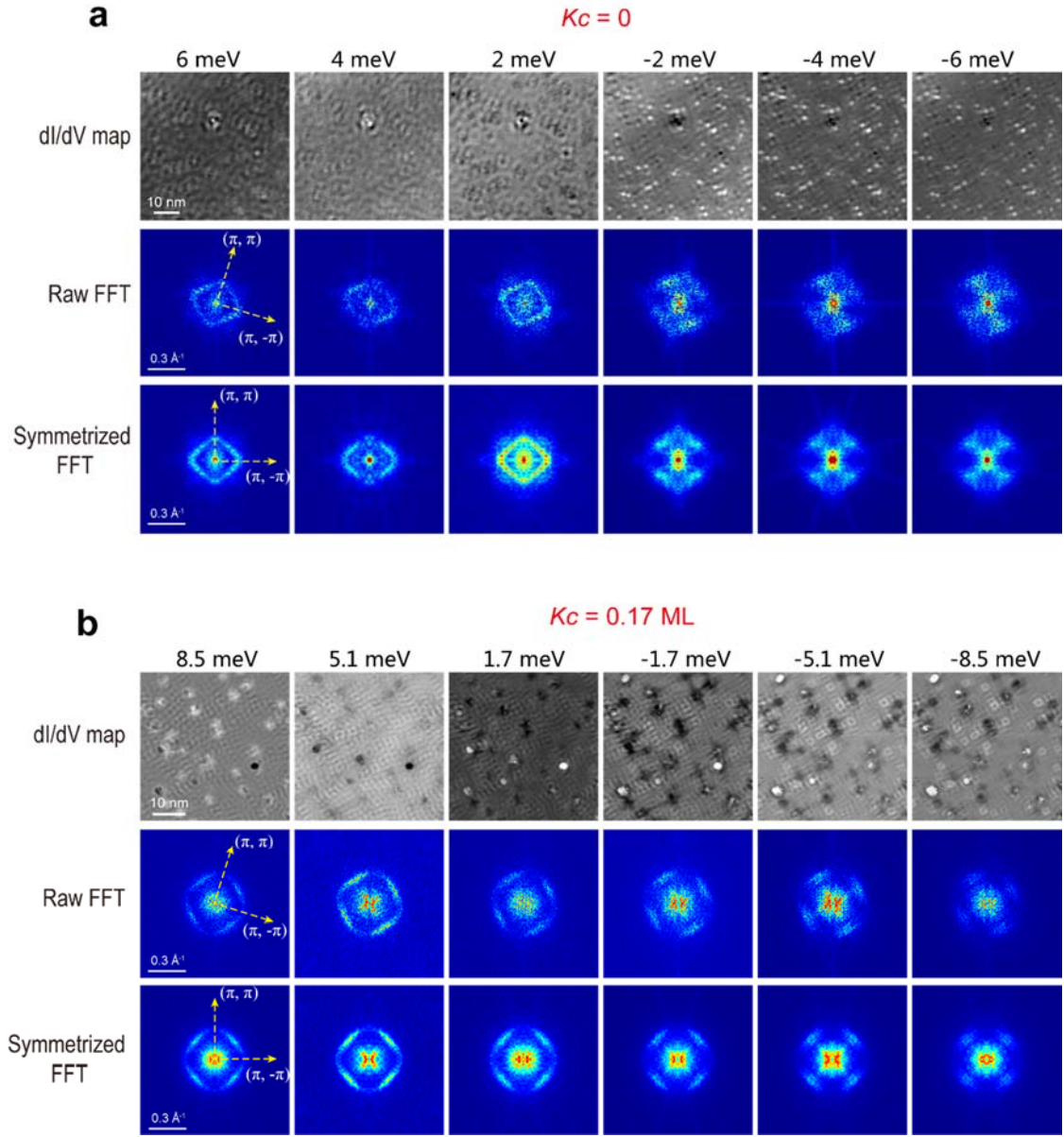

**Figure S9 |** (**a**) Additional dI/dV maps, raw FFTs and symmetrized FFTs taken on RbFe$_2$As$_2$ with Kc =0 at T = 4.5K. All dI/dV maps are taken at the setpoint of V$_b$ = 10mV, I = 200pA, ΔV = 1mV. (**b**) Additional dI/dV maps, raw FFTs and symmetrized FFTs taken on RbFe$_2$As$_2$ with Kc =0.17 ML at T = 4.5K. All dI/dV maps are taken at the setpoint of V$_b$ = 10mV, I = 200pA, ΔV = 1mV. Each map has 250 × 250 pixels.

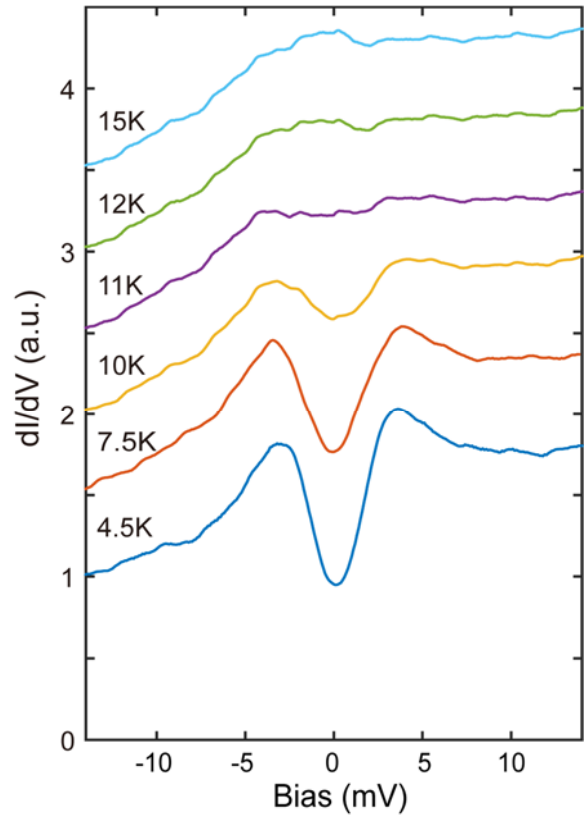

**Figure S10 |** Temperature dependence of the low energy tunneling gap observed at Kc = 0.17 ML ($V_b$ = 15mV, I = 300pA, $\Delta V$ = 1mV for all the spectra). The gap closed at about 12K.